\documentstyle[amsfonts,prl,aps,multicol]{revtex}
\input psfig
\pssilent
\begin{document}
\draft

\title{Dynamical Initial Conditions in Quantum Cosmology}
\author{Martin Bojowald\cite{Email}}
\address{Center for Gravitational Physics and Geometry, The
  Pennsylvania State University,\\ 104 Davey Lab, University Park, PA
  16802, USA}

\maketitle

\begin{abstract}
  Loop quantum cosmology is shown to provide both the dynamical law
  and initial conditions for the wave function of a universe by one
  discrete evolution equation. Accompanied by the condition that
  semiclassical behavior is obtained at large volume, a unique wave
  function is predicted.
\end{abstract}

\vspace{-4.5cm} 
\begin{flushright}
CGPG--01/4--3  \\
gr-qc/0104072 \\
\end{flushright}
\vspace{3.5cm}


\begin{multicols}{2}

Traditionally, physical systems are modeled mathematically by
providing laws governing the dynamical behavior and specifying initial
(or boundary) conditions. The latter select a particular solution to
the laws, but usually all of them are allowed and describe the system
under different conditions. However, in cosmology the situation is
different: there is only one universe, and therefore only one fixed
set of initial conditions can lead to the physically realized
situation. In this context, the big bang singularity is regarded as
the point of ``creation'' of the universe at which initial conditions
(or equivalent restricting requirements) have to be imposed. Since
gravity is strong at that stage and classical general relativity
breaks down (signaled by the appearance of a classical singularity), a
quantization of the gravitational field is needed bringing us in the
realm of quantum cosmology.

The standard approach to quantum cosmology consists in quantizing a
minisuperspace model which is obtained by specifying symmetry
conditions, usually homogeneity and isotropy, for the allowed metrics
in space-like slices of a universe. This reduces the infinitely many
degrees of freedom of general relativity to finitely many ones
allowing standard quantum mechanical methods \cite{DeWitt,Misner}. Due
to general covariance the dynamical law is provided by a constraint
equation which takes the form of a second order differential
equation---the Wheeler--DeWitt equation---for the wave function
$\psi(a,\phi)$ depending on the scale factor $a>0$ (conventionally
used as internal time) and matter degrees of freedom collectively
denoted by $\phi$. However, the classical singularity remains, and no
initial conditions are provided by the formalism which leads at least
to a two-parameter family (not counting matter degrees of freedom) of
solutions and not a unique (up to norm) one. The original hopes
\cite{DeWitt} that there might be a unique solution to the constraint
equation are not realized.

To address this issue, proposals have been developed by several
authors. However, these proposals have considerable arbitrariness
since they are driven primarily by the authors' intuition as to how
the classical singularity might be smoothed out by quantum
gravity. Most well-known are the ``no-boundary'' proposal
\cite{nobound} and the ``tunneling'' proposal \cite{tunneling} which
both describe the ``creation'' of a universe at the place of the
classical singularity. In all those approaches matter is regarded as
being irrelevant in the early stages, and so the wave function $\psi$
is assumed (implicitly or explicitly \cite{SIC}) to be independent of
the variables $\phi$ for small $a$; this is already an initial
condition which strongly restricts the $\phi$-dependence of $\psi$. We
will take the same point of view concerning matter degrees of freedom
here, which we regard as being justified thanks to the dominance of
gravity in early stages of the evolution.

But still, there is a two-parameter family of solutions $\psi(a)$ from
which one parameter has to be fixed (since the norm is
irrelevant). This not only influences the wave function close to the
singularity, but also its late time behavior because it selects a
particular linear combination of the expanding and contracting
components in a WKB-approximation. However, as an initial condition it
is specified at the classical singularity (e.g., by fixing the value
$\psi(0)$ \cite{DeWitt,Konto} or by introducing an ad hoc ``Planck
potential'' \cite{SIC}), and thus involves Planck scale physics for
which we need a full quantum theory of gravity.

One candidate for a quantization of general relativity is quantum
geometry (see e.g.\ \cite{Nonpert,Rov:Loops}) which predicts discrete
eigenvalues of geometrical operators like area and volume
\cite{AreaVol,Area,Vol2}. A symmetry reduction \cite{SymmRed} of the
quantized (kinematical) theory to cosmological models leads to loop
quantum cosmology \cite{cosmoI}, in which the discreteness of the
volume is preserved \cite{cosmoII}. All techniques used in this
framework of quantum cosmology are very close to those of {\em full\/}
quantum gravity, in contrast to standard quantum cosmology which is
based on a {\em classical\/} symmetry reduction to a simple mechanical
system and subsequent quantization. Hence, the results of loop quantum
cosmology should be more reliable, in particular close to the
classical singularity where the two approaches show the largest
differences. In fact, loop quantum cosmology has a {\em discrete\/}
evolution equation \cite{cosmoIII,cosmoIV} which replaces the
Wheeler--DeWitt equation and is {\em singularity-free\/}
\cite{Sing,InvScale,IsoCosmo}. The fact that the Hamiltonian
constraint operator of loop quantum cosmology \cite{cosmoIII} is very
close to that of the full theory \cite{QSDI} gives rise to the hope
that the results of \cite{Sing} can be extended to less symmetric
models.

Loop and standard quantum cosmology deviate most when applied right at
the classical singularity. In this letter we will show that the
particular form of the evolution equation of loop quantum cosmology,
applied at vanishing scale factor, leads to a consistency condition
for the initial data. In this way the evolution equation provides both
the dynamical law and initial conditions: {\em dynamics dictates the
initial conditions}. Accompanied by a classicality condition for the
solutions, a unique (up to norm) wave function is predicted.

\paragraph*{Isotropic loop quantum cosmology.}

In the triad representation of isotropic loop quantum cosmology
\cite{IsoCosmo} the scale factor $a\in{\Bbb R}^+$ is replaced by a
discrete label $n\in{\Bbb Z}$ which parameterizes eigenvalues of the
triad operator. An orthonormal basis of the kinematical Hilbert space
is given by quantum states $|n\rangle$ labeled by the triad eigenvalue
$n$ which also determines volume eigenvalues: $\hat{V}|n\rangle=
V_{(|n|-1)/2} |n\rangle$ with
\begin{equation}\label{Vj}
 V_j=(\gamma l_{\rm P}^2)^{\frac{3}{2}}
 \sqrt{\case{1}{27}j(j+\case{1}{2})(j+1)}
\end{equation}
($\gamma\in{\Bbb R}^+$, which is of order one, is the Barbero--Immirzi
parameter labeling inequivalent representations of the classical
Poisson algebra, and $l_{\rm P}=\sqrt{\kappa\hbar}$ with $\kappa=8\pi
G$ is the Planck length). The volume operator has eigenvalue zero with
threefold degeneracy (for the states $|\pm 1\rangle$ and $|0\rangle$),
but only one of them, $|0\rangle$, has degenerate triad and so
corresponds to the classically singular state. The wave function
$\psi(a,\phi)$ of standard quantum cosmology is replaced by the
coefficients $s_n(\phi)$ of a state $|s\rangle=\sum_n
s_n|n\rangle$. For large $|n|$ the correspondence between $a$ and $n$
is $|n(a)|\sim 6a^2\gamma^{-1} l_{\rm P}^{-2}$ which follows from the
volume spectrum ($|n|=2j+1$).

The Hamiltonian constraint equation for spatially flat models takes
the form of a discrete evolution equation (see \cite{IsoCosmo} for the
case of models with positive spatial curvature):
\begin{eqnarray}\label{Lor}
&&\case{1}{4} (1+\gamma^{-2}) A_n^{(8)} s_{n+8}(\phi)- A_n^{(4)}
s_{n+4}(\phi) - 2 A_n^{(0)} s_n(\phi)\nonumber\\
&& - A_n^{(-4)} s_{n-4}(\phi) +\case{1}{4} (1+\gamma^{-2}) A_n^{(-8)}
 s_{n-8}(\phi) \nonumber\\
&&\quad  = -\case{1}{3}\gamma\kappa\l_{\rm P}^2
 \,\hat{H}_{\phi}(n)\, s_n(\phi)
\end{eqnarray}
with
\begin{eqnarray}
 A_n^{(\pm 8)} &:=&
  \left(V_{|n\pm 8|/2}- V_{|n\pm 8|/2-1}\right)
  k_{n\pm 8}^{\pm}k_{n\pm 4}^{\pm} \\
 A_n^{(\pm 4)} &:=&
  \left(V_{|n\pm 4|/2}- V_{|n\pm 4|/2-1}\right)\\
 A_n^{(0)} &:=&
  \left(V_{|n|/2}-
   V_{|n|/2-1}\right) \nonumber\\
 && \times\left(\case{1}{8} (1+\gamma^{-2})
   (k_n^-k_{n+4}^++ k_n^+k_{n-4}^-) -1\right)
\end{eqnarray}
where $V_j$ are the eigenvalues (\ref{Vj}) of the volume operator with
$V_{-1}=0$, and the coefficients $k_n^{\pm}$ can be chosen to be
non-vanishing by a suitable ordering of the extrinsic curvature
operator and are approximately ${\rm sgn} (n)$ for large $|n|$ (see
\cite{IsoCosmo} for explicit expressions in terms of the volume
eigenvalues). Here we introduced a matter Hamiltonian $\hat{H}_{\phi}$
whose particular form is irrelevant. It only matters that it acts
diagonally in the triad degrees of freedom which is always the case in
the absence of curvature couplings.

\paragraph*{Pre-Classicality.}

Compared to the standard Wheeler--DeWitt equation of second order the
discrete evolution equation is of order sixteen. So the problem of a
unique solution seems to be more severe at first sight, but many of
the additional solutions can easily be seen to not correspond to a
semiclassical solution. Let us call a wave function $s_n$ {\em
pre-classical\/} if and only if, at large volume ($n\gg 1$), it is not
strongly varying at the Planck scale (increasing the large label $n$
by one), although it may oscillate on much larger scales (increasing
$n$ by a macroscopic amount). Note that a difference equation with
{\em fixed step size}, as is always the case here, may have solutions
which are very different from those of an approximating differential
equation even though all solutions of the differential equation are
well approximated when the step size goes to zero \cite{SIAM}. A
common possibility is a solution with alternating sign between
successive $n$, which cannot correspond to a continuous solution of a
differential equation. Due to instabilities there can also be
solutions with exponentially increasing absolute value, even in
regimes where the solutions of the differential equation are purely
oscillating (i.e.\ in the classically allowed range in a WKB
approximation).  A precise formulation of the phrase ``not strongly
varying'' can be given in the following way. Note first that the
Barbero--Immirzi parameter $\gamma$ enters $a(n)=\gamma l_{\rm
P}^2n/6$, which is used here as internal time. Although the physical
value of $\gamma$ is fixed and of order one \cite{ABCK:LoopEntro}, we
can use the $\gamma\to0$ limit, together with $n\to\infty$ such that
$a(n)$ is fixed, to decide whether a wave function is
pre-classical. In this limit the difference $a(n+1)-a(n)$ becomes
infinitesimal implying a continuum limit. A wave function $s_n$ is
pre-classical if and only if its limit $\gamma\to0$, $n\to\infty$
exists, providing a rigorous check of the pre-classicality
condition. Note that $\kappa$ and $\hbar$, and so $l_{\rm P}$, are
fixed in this limit and we are still dealing with quantum
cosmology. In fact, standard quantum cosmology can be shown to be the
above limit of loop quantum cosmology.

\begin{figure}
 \centerline{\psfig{figure=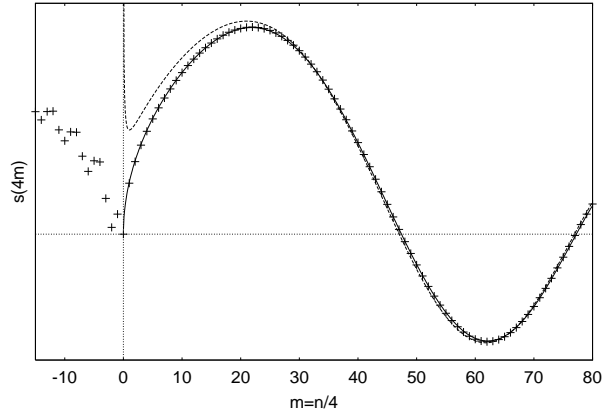,width=3.25in}} \caption{The
 unique solution ($+$) of the Hamiltonian constraint (\ref{Lor}), when
 the Ricci curvature is only due to a positive cosmological constant
 $\lambda:=l_{\rm P}^2 \Lambda= 2\cdot 10^{-4}$ ($\gamma=1$), which is
 pre-classical for large positive $n$. Evolving backwards through
 $n=0$, the wave function picks up a wildly oscillating component and
 is no longer exactly pre-classical at negative $n$. The standard
 quantum cosmology wave function $\psi(a)$, subject to $(l_{\rm P}^4
 (4a)^{-1}{\rm d}/{\rm d}a(a^{-1}{\rm d}/{\rm d}a)+ 3\lambda a^2l_{\rm
 P}^{-2})\sqrt{a}\psi(a)=0$ in the ordering corresponding to
 (\ref{Lor}), is given by $\psi(a)=a^{-\frac{1}{2}}(A\, {\rm Ai}(
 -(3\lambda)^{\frac{1}{3}} a^2l_{\rm P}^{-2})+B\, {\rm Bi}(
 -(3\lambda)^{\frac{1}{3}} a^2l_{\rm P}^{-2}))$ in terms of Airy
 functions. Wave functions $\psi(a)$ for two choices of the parameters
 $A$ and $B$ are shown: the continuous line is the unique (up to norm)
 choice fulfilling DeWitt's $\psi(0)=0$, which in this case is in good
 agreement with $s_n$ at positive $n$, whereas any other choice leads
 to a diverging wave function (dashed line).  \label{deSitter}}
\end{figure}

Our condition picks out only those solutions which are oscillatory on
large scales but almost constant on the Planck scale. Since this is a
pre-requisite for a subsequent WKB-approximation, we call it
pre-classicality.  Whenever it is fulfilled, a discrete wave function
$s_n(\phi)$ can be approximated at large $n$ by a standard continuous
wave function $\psi(a):=s_{n(a)}$ with $n(a)=6a^2 \gamma^{-1}l_{\rm
P}^{-2}$ as above, which approximately solves the standard
Wheeler--DeWitt equation up to corrections of order
$\sqrt{\gamma}l_{\rm P}/a$ \cite{Sing,IsoCosmo}. Thus, standard
quantum cosmology is realized only as an approximation valid at large
volume where the discreteness of quantum geometry is irrelevant (see
Fig.\ \ref{deSitter}).

Since the Wheeler--DeWitt equation is of second order and so has two
independent solutions, there can be at most two independent
pre-classical solutions $s^{\pm}$ of our discrete evolution equations,
such that any pre-classical solution can be written as $s=as^++bs^-$
with $a,b\in{\Bbb C}$.

To demonstrate this explicitly, we introduce
\begin{eqnarray*}
 t_m &:=& \gamma^{-1}l_{\rm P}^{-2}(V_{2|m|}-V_{2|m|-1})s_{4m}\\
 P(m) &:=& \case{1}{3}\gamma\kappa l_{\rm P}^2H_{\phi}(m)
 (V_{2|m|}-V_{2|m|-1})^{-1}\,,
\end{eqnarray*}
using the expectation value $H_{\phi}(n)$ of $\hat{H}_{\phi}(n)$ in a
matter state, such that for $|n|\gg1$, where $k_n^{\pm}\sim{\rm
sgn}(n)$, the evolution equation (\ref{Lor}) takes the form
\begin{eqnarray}
 &&\case{1}{4}(1+\gamma^{-2}) t_{m+2}-t_{m+1}+
 (\case{1}{2}(3-\gamma^{-2})+ P(m))t_m\nonumber\\
 &&-t_{m-1}+ \case{1}{4}(1+\gamma^{-2}) t_{m-2}=0\,. \label{LorSimp}
\end{eqnarray}
In a classical regime $|m|$ is large and $P(m)\sim
\frac{2}{3}\kappa H_{\phi}/a$ is approximately constant on a range
small compared to $|m|$. In this case we have a linear difference
equation with constant coefficients whose solutions can be found by an
ansatz $t_m\propto e^{im\theta}$ with $\theta\in{\Bbb C}$ which in
(\ref{LorSimp}) yields the quadratic equation
\[
 (1+\gamma^{-2})\cos^2\theta-2\cos\theta+1-\gamma^{-2}+
 P=0
\]
which has solutions
\[
 \cos\theta=(1+\gamma^{-2})^{-1}\left(1\pm\sqrt{\gamma^{-4}-
 (1+\gamma^{-2})P}\right)
\]
being real with modulus smaller than one such that $\theta$ is real when
$\gamma$ is of the order one and $P$ is small.

If the matter does not contribute a Planck size energy, $P$ is small
and we have $\cos\theta_0= 1-\epsilon+O(\epsilon^2)$ or $\cos\theta_1=
(1+\gamma^{-2})^{-1} (1-\gamma^{-2})+ \epsilon+O(\epsilon^2)$ with
$0<\epsilon:=\frac{1}{2}\gamma^2P\ll 1$. The first possibility,
expanding $\cos\theta_0= 1-\frac{1}{2}\theta_0^2+O(\theta_0^4)$, leads
to two solutions $\theta_0=\gamma\sqrt{P}+O(P)$ and $-\theta_0$ with
$|\theta_0|\ll 1$ both of which imply pre-classical $t_m^{\pm}=e^{\pm
im\theta_0}$. Because $\gamma$ is not large compared to one, the
second possibility $\cos\theta_1$ leads to $\theta_1$ which violates
pre-classicality (e.g., for $\gamma=1$ we have $\theta_1=\pm\pi/2$ and
$t_m\propto (\pm i)^m$).

All 16 independent solutions $t_{n/4}=\gamma^{-1}l_{\rm P}^{-2}
(V_{|n|/2}-V_{|n|/2-1})s_n$ of (\ref{Lor}) can be obtained as
\begin{eqnarray*}
 t_{n/4} &=& e^{\pm
 in\theta_0/4},\: e^{\pm in\theta_1/4},\\
 && (-1)^n e^{\pm in\theta_0/4},\:
  (-1)^n e^{\pm in\theta_1/4},\\
 && \sigma^n e^{\pm in\theta_0/4} \mbox{
 or } \sigma^n e^{\pm in\theta_1/4}
\end{eqnarray*}
where $\sigma$ can be $+i$ or $-i$. Obviously, only the first two are
pre-classical. The definition using $\gamma\to0$ ($a$ finite) is
applied as follows: with $n=6a^2\gamma^{-1}l_{\rm P}^{-2}$ we have
\begin{eqnarray*}
 \lim_{\gamma\to0} \gamma^{-1} l_{\rm P}^{-2} (V_{|n|/2}-
 V_{|n|/2-1})= a/2\,,\\
 \lim_{\gamma\to0}e^{\pm in\theta_0/4}= \exp\left(\pm \case{3}{2} i\sqrt{P}
 a^2/l_{\rm P}^2\right)\,, 
\end{eqnarray*}
whereas $\lim_{\gamma\to0} \theta_1= \pi$ and so the
limit for $\gamma\to0$, $n\to\infty$ does not exist for the remaining
14 solutions.

Thus, by using the pre-classicality requirement, which is a
prerequisite for any semiclassical analysis, we arrive at the same
situation as in standard quantum cosmology: there are two
independent solutions from which we have to select a linear
combination up to norm.

\paragraph*{Dynamical Initial Conditions.}

Up to now we considered only the semiclassical regime, but there is an
additional feature of loop quantum cosmology {\em which emerges right
at the classical singularity}, deeply in the Planck regime where the
approximation by standard quantum cosmology breaks down: the highest
order (or lowest order when we evolve backwards) coefficient vanishes
when we try to determine $s_0$. At first sight, it seems that this is
a breakdown of the evolution similar to the classical
situation. However, as demonstrated in \cite{Sing,IsoCosmo}, this is
not the case in the particular factor ordering of the constraint
chosen above because $s_0$ completely drops out of the evolution
equation.  (This observation depends crucially on the fact that
$\hat{H}_{\phi}s_0(\phi)=0$ which is always true in quantum geometry
\cite{InvScale,IsoCosmo} but would be impossible without space-time
discreteness.)

Instead of determining $s_0$ the evolution equation leads to a {\em
consistency condition for the initial data}: starting from a general
pre-classical solution $s_n=as_n^++bs_n^-$ for large $n$ and evolving
backwards, we eventually arrive at a point where we have to apply
(\ref{Lor}) for $n=8$. At this value of $n$ the lowest order
coefficient $A_8^{(-8)}$ vanishes as noted above causing $s_0$ to drop
out. Since by assumption we have already determined all $s_n$ for
$n>0$ (for which there is no vanishing coefficient in the evolution
equation), the would-be equation for $s_0$ leads to a further
condition for higher $s_n$ ($s_4$, $s_8$, $s_{12}$ and $s_{16}$) which
upon inserting the general pre-classical solution $s_n=as_n^++bs_n^-$
implicitly yields a linear relation between the two free parameters
$a$ and $b$. {\em This leaves us with a unique solution\/} (up to
norm).

\paragraph*{Conclusions.}

We have shown that loop quantum cosmology implies a discrete evolution
equation which uniquely determines a state (up to norm) behaving
semiclassically at large volume. It is important to adapt the standard
condition for semiclassicality in a WKB approximation taking the
discreteness of time into account. This leads already to a strong
reduction of the allowed solutions, but the crucial condition for the
uniqueness arises only from the particular structure of the evolution
equation in quantum geometry. We remark that in general it is only
possible to require pre-classicality at one connected domain of large
volume. If one evolves through a classical singularity, the wave
function may pick up components which oscillate at the Planck scale
(see Fig.\ \ref{deSitter}). The precise form of these oscillations
depends on factor ordering ambiguities (in the coefficients
$k_n^{\pm}$ entering the constraint) and the use of the Lorentzian
(versus Euclidean) theory.

Such a unique wave function generally differs from those obtained with
boundary proposals of standard quantum cosmology. By choosing a real
prefactor it is always real (for flat spatial slices the evolution
equation has real coefficients; this no longer holds true for
spatially curved models) and so cannot coincide with the ``tunneling''
wave function \cite{tunneling}. While the ``no-boundary'' proposal
\cite{nobound} also leads to a real wave function, it is imposed on
the standard Wheeler--DeWitt equation at the Planck scale where large
deviations to loop quantum cosmology occur. Thus, in general its wave
function of a universe is different from the unique pre-classical
solution found here. The consistency condition for the initial data in
loop quantum cosmology may be expressed as $s_0=0$ which is
reminiscent of DeWitt's $\psi(0)=0$ \cite{DeWitt} (to achieve this, an
ad hoc Planck potential has been introduced in \cite{SIC}). However,
since these two conditions are imposed on completely different
evolution equations, the selected solutions in general differ. As
Fig.\ \ref{deSitter} shows, there may be a good coincidence in certain
models, but only if the curvature is small at all times which can
happen only in the absence of matter.

Contrary to all other proposals for boundary conditions in quantum
cosmology, our {\em dynamical initial conditions\/} are not chosen to
fulfill an a priori intuition about the ``creation'' of a universe but
derived from the evolution equation which, in turn, is derived from
quantum geometry, a candidate for a complete theory of quantum
gravity. Therefore, one equation provides both the dynamical law and
initial conditions. As we have seen, the critical condition, which
crucially depends on quantum geometry, emerges from evaluating the
evolution equation at the state which corresponds to the classical
singularity. So in contrast to the classical situation where a
singularity leads to unpredictability, in quantum geometry the regime
of the classical singularity fixes ambiguities in the wave function of
a universe.

\paragraph*{Acknowledgements.}

The author is grateful to A.\ Ashtekar for discussions and a careful
reading of the manuscript.
This work was supported in part by NSF grant PHY00-90091 and the Eberly
research funds of Penn State.

\end{multicols}
\end{document}